# Comparison of Machine Learning Classifiers to Predict Patient Survival and Genetics of GBM: Towards a Standardized Model for Clinical Implementation

Luca Pasquini[1,2], Antonio Napolitano[3], Martina Lucignani[3], Emanuela Tagliente[3], Francesco Dellepiane[2], Maria Camilla Rossi-Espagnet[2,4], Matteo, Ritrovato[5], Antonello Vidiri[6], Veronica Villani[7], Giulio Ranazzi[8], Antonella Stoppacciaro[8], Andrea Romano[2], Alberto Di Napoli[2,9], Alessandro Bozzao[2]

[1]Neuroradiology Unit, Radiology Department, Memorial Sloan Kettering Cancer Center, 1275 York Ave, New York, NY 10065, USA.
[2]Neuroradiology Unit, NESMOS Department, Sant'Andrea Hospital, La Sapienza University, Via di Grottarossa 1035, Rome 00189, Italy.
[3]Medical Physics Department, Bambino Gesù Children's Hospital, IRCCS, Piazza di Sant'Onofrio, 4, Rome 00165, Italy.
[4]Neuroradiology Unit, Imaging Department, Bambino Gesù Children's Hospital, IRCCS, Piazza di Sant'Onofrio, 4, Rome 00165, Italy.
[5]Unit of HTA, Biomedical Technology Risk Manager, Bambino Gesù Children's Hospital, IRCCS, Piazza di Sant'Onofrio, 4, Rome 00165, Italy.
[6]Radiology and Diagnostic Imaging Department, Regina Elena National Cancer Institute, IRCCS, Via Elio Chianesi 53, Rome 00144, Italy.
[7]Neuro-Oncology Unit, Regina Elena National Cancer Institute, IRCCS, Via Elio Chianesi 53, Rome 00144, Italy.
[8]Department of Clinical and Molecular Medicine, Surgical Pathology Units, Sant'Andrea Hospital, La Sapienza University, Via di Grottarossa 1035, Rome 00189, Italy.
[9]Radiology Department, Castelli Romani Hospital, Via Nettunense Km 11.5, Ariccia 00040, Rome, Italy.

*Correspondence:*
Luca Pasquini
Email: lucapasquini3@gmail.com ; Tel.: +1-883-980-2634; +39-06-3377-5225

**Abstract:** Radiomic models have been shown to outperform clinical data for outcome prediction in glioblastoma (GBM). However, clinical implementation is limited by lack of parameters standardization. We aimed to compare nine machine learning classifiers, with different optimization parameters, to predict overall survival (OS), isocitrate dehydrogenase (IDH) mutation, O-6-methylguanine-DNA-methyltransferase (MGMT) promoter methylation, epidermal growth factor receptor (EGFR) VII amplification and Ki-67 expression in GBM patients, based on radiomic features from conventional and advanced MR. 156 adult patients with pathologic diagnosis of GBM were included. Three tumoral regions were analyzed: contrast-enhancing tumor, necrosis and non-enhancing tumor, selected by manual segmentation. Radiomic features were extracted with a custom version of Pyradiomics, and selected through Boruta algorithm. A Grid Search algorithm was applied when computing 4 times K-fold cross validation (K=10) to get the highest mean and lowest spread of accuracy. Once optimal parameters were identified, model performances were assessed in terms of Area Under The Curve-Receiver Operating Characteristics (AUC-ROC). Metaheuristic and ensemble classifiers showed the best performance across tasks. xGB obtained maximum accuracy for OS (74.5%), AB for IDH mutation (88%), MGMT methylation (71,7%), Ki-67 expression (86,6%), and EGFR amplification (81,6%). Best performing features shed light on possible correlations between MR and tumor histology.

## 1. Introduction

In recent years, artificial intelligence (AI) applications in biomedical imaging have grown exponentially. The conversion of radiologic images in mineable data and their analysis with AI techniques to support medical decisions is defined 'radiomics' (1). Biomedical images intrinsic parameters can reflect tissue structure, molecular data and patient outcome, providing important information for patient care through quantitative image analyses (1,2).

GBM is considered the most frequent and lethal primary malignant tumor for adults, with an estimated incidence rate of 3.19 per 100,000 persons in the United States, and a median age of 64 years (3). Despite combined radio-chemotherapy, the OS of patients with GBM is dismally poor (4). Genetic alterations such as IDH mutation, MGMT promoter methylation, EGFR VII amplification may influence patient outcome, with effects on survival, progression, and treatment response (4,5).

Radiomic-based predictive models have been shown to outperform clinical models based on patient age, Karnofsky performance scale, type of surgical resection, genetic alterations, for outcome prediction (6,7), with 36% improvement for progression free survival (PFS) and 37% for OS prediction (6). Recent studies in patients affected by GBM displayed several high-performance radiomic models for predicting OS, PFS, molecular subtypes of GBM, as well as genetic alterations critical for the clinical practice (8–11). Despite these promising results, clinical implementation is extremely limited due to wide variations of model performances (12,13), and controversial findings. For example, a recent study on 152 patients with GBM concluded that Magnetic Resonance (MR) imaging features were not adequate for providing reliable and clinically meaningful predictions through machine learning (ML) classification models (14).

Variability in model performance may depend on parameters optimization. Radiomic workflows comprehend multiple steps requiring parameter choice: tumor segmentation on radiologic images, feature extraction and selection, training, testing and validation of the AI model, performance evaluation (15,16). The lack of radiomic parameters standardization might limit results generalizability across studies. A possible solution for this limitation is to compare multiple ML algorithms in the same population for different tasks. In fact, the classification method was shown to be the dominant source of performance variation in radiomic analyses (17).

Comparative radiomic studies have been rarely performed in the literature, with suboptimal results. Parmar et al. examined fourteen feature selection methods and twelve classification methods for predicting OS from pretreatment computed tomography images of 464 lung cancer patients, showing the best performance for a random forest classifier (AUC 0.65) (17). Another study compared eight ML algorithms for predicting OS from conventional MR images of 163 patients with GBM. The highest performance was achieved by an ensemble learning model, with poor results in terms of overall accuracy (57.8%) (18).

The aim of our study was to compare nine ML classifiers to predict OS, IDH mutation, MGMT promoter methylation, EGFR VII amplification and Ki-67 expression in the same population of patients affected by GBM, based on features extracted from conventional and advanced MR images. We tested different optimization parameters and reported classification results to provide a comprehensive view on model performance for relevant tasks in the clinical practice.

## 2. Materials and Methods

*2.1 Subjects*

This retrospective observational study was conducted in agreement with the Helsinki declaration and was approved by the institutional IRB (protocol number: 19 SA_2020). We selected patients with a diagnosis of Glioblastoma (GBM), who underwent preoperative MRI from March 2005 to May 2019. Data was collected from two institutions: Sant'Andrea Hospital La Sapienza University of Rome (Institution 1) on a 1.5T scanner (Magnetom Sonata, Siemens, Erlangen, Germany), and Regina Elena Institute of Rome (Institution 2) on a 3T system (Discovery MR 750w, GE Healthcare, Milwaukee, WI, USA). We enrolled patients fulfilling the following inclusion criteria: histopathological diagnosis of GBM, MRI acquisitions in the preoperative phase with at least one among structural, diffusion or perfusion techniques. Exclusion criteria were motion artefacts or other causes of sub-optimal images, loss of patients' information during follow-up.

All patients received postoperative focal radiotherapy (RT) plus concomitant daily temozolomide (TMZ), followed by adjuvant TMZ therapy, with the same treatment protocol. RT started within 4 weeks of surgery and consisted of fractionated focal irradiation at a dose of 60 Gy, delivered in 30 fractions of 2 Gy over 6 weeks. Concomitant chemotherapy consisted of TMZ in a dose of 75 mg/m2 administered 7 days/week from the first day of RT. Adjuvant TMZ therapy began 4 weeks after the end of RT and was delivered for 5 days every 28 days, up to 12 cycles. The dose was 150 mg/m2 for the first cycle and was increased to 200 mg/m2 for the second one.

*2.2 Histopathological Analysis*

The specimens were fixed in 10% formaldehyde and processed for paraffin embedding. Two μm thick sections were mounted and stained with hematoxylin and eosin. Histopathological examination, typing and grading were performed identifying at least three of the following features in astrocytic tumors: mitotic figures, cellular atypia, microvascular proliferation and/or necrosis, according to the last edition of the World Health Organization classification of CNS tumors.

*2.3 Immunohistochemistry*

Immunohistochemistry was performed using Dako Envision Flex system. The immunostaining patterns of EGFR were evaluated considering both cellular and tissue distribution. The number of immunopositive cells in ten high power (40x) areas were counted and the percentage of immunopositive cells were estimated. The ratio of positive cells/total number of cells was calculated for each field. The mean value of the ten fields obtained from a section was considered as the estimated percentage of immunoreactivity assigned to the tumor sample. For the evaluation of IDH-1 mutation, IDH-1 R132H antibody was used. The test was defined as positive if a focal or diffuse immunopositivity was detected and negative if no tumor cells showed immunopositivity. Negative cases were then analyzed for IDH-1/2 mutations by directly sequencing the exon 4 of *IDH1* gene including codon 132, and a fragment of 219 bp in length spanning the catalytic domain of *IDH2* including codon 172 following PCR amplification of genomic DNA using respectively the primers *IDH-1*: Forward, 5′-CGG TCT TCA GAG AAG CCA TT-3′, Reverse, 5′- ATT CTT ATC TTT TGG TAT CTA CAC C-3′, *IDH-2*: forward 5'-CAAGCTGAAGAAGATGTGGAA-3', reverse 5′ CAGAGACAAGAGGATGGCTA-3′. All sequence reactions were carried out using the GenomeLab DTCS quick-start kit (Beckman Coulter, Fullerton, CA, USA). The reactions were carried out in an automated DNA analyzer (CEQ 8000; Beckman Coulter). All sections were immunostained with Ki-67 antibody. The positivity for Ki67 was determined by counting at least 1000 tumor cells in a homogeneously stained area and then expressed in percentage.

*2.4 MGMT methylation testing*

We used EntroGen's MGMT Methylation Detection Kit (MSPCR, Cat. No. MGMT-RT44), a semi-quantitative real-time PCR-based essay for detection of MGMT promoter methylation within the DMR2 locus, distinguishing between methylated and non-methylated cytosines. Its target region starts at chr10:131265513 (hg19 genome build) in the MGMT promoter region and covers CpG sites 75-86. The detection of the amplification product was done by using fluorescent hydrolysis fraction. The procedure involves the following steps: 1) isolation of DNA from tumor biopsies, paraffin embedded sections; 2) bisulfite treatment of the isolated DNA using the EZ DNA methylation-Lightning Kit (Zymo Research, CATD5030); 3) amplification of treated DNA using the provided reagents in the MGMT Promoter methylation Detection kit; 4) data analysis and interpretation using the real time PCR software.

*2.5 MR image acquisition*

Acquired MR sequences included MPRAGE, FLAIR, T1-weighted, T2-weigthed, diffusion weighted images (DWI), with apparent diffusion coefficient (ADC) map reconstruction, and perfusion weighted images (PWI) with dynamic susceptibility contrast (DSC) technique.

Patients acquired with the 1.5T scanner underwent the following protocol: axial T1-weighted spin echo (TR/TE, 600/12 ms; ST 4 mm; FA 150°; matrix 512 x 512), axial T2-weighted fast spin echo (TR/TE 2920/107; ST 4 mm; FA 180°; matrix 512 x 512), axial FLAIR (TR/TE 10000/126 ms; ST 4 mm, IT 2500 ms; FA 150°; matrix 512 x 512). DWI (TR/TE 3000/84 ms; ST 5 mm; FA 90°; matrix, 256 x 256) were acquired with three levels of diffusion sensitization (b-values 0, 500 and 1000). DSC images acquired during contrast injection (DOTAREM.; dose 0.1 mmol/kg, injection rate 4 ml/s) followed by a 20-ml saline flush, based on T2*- weighted gradient-echo echo-planar sequence (TR/TE 1490/40 ms; flip angle 90°; FOV 230 x 230 mm; matrix 128 x 128, 14 sections of 5 mm thickness, 50 volumes). MPRAGE (TR/TE 1840/4.4 ms; ST 1 mm; IT 1100 ms; FA 15°; matrix 256 x 256) after administration of contrast.

Patients acquired with the 3T scanner underwent the same protocol, with the following sequences: axial T1-weighted spin echo (TR/TE, 600/6.4 ms; ST 3 mm; FA 138°; matrix 512x512), axial T2-weighted fast spin echo (TR/TE 8694/148; ST 3 mm; FA 160°; matrix 512 x 512), axial FLAIR (TR/TE 7002/142 ms; ST 1 mm, IT 1892 ms; matrix 512 x 512). DWI (TR/TE 6500/88; ST 4 mm; FA 90°; matrix 256 x 256) were acquired with three levels of diffusion sensitization (b-values 0, 500 and 1000). DSC images acquired during contrast injection (DOTAREM.; dose 0.1 mmol/kg, injection rate 4 ml/s) followed by a 20-ml saline flush, based on T2*- weighted gradient-echo echo-planar sequence (TR/TE 2000/16 ms; flip angle 60°; FOV 230 x 230 mm; matrix 128 x 128, 20 sections of 4 mm thickness, 45 volumes). BRAVO (TR/TE 8.524/3.3 ms, ST 1 mm, IT 450 ms, FA 12°, matrix 256 x 256) after administration of contrast.

Perfusion parametric maps were obtained through a dedicated software package OleaSphere software version 3.0 (Olea Medical, La Ciotat, France). A relative cerebral blood volume (rCBV) map was generated by using an established tracer kinetic model applied to the first-pass data (19). As previously described (20), the dynamic curves were mathematically corrected to reduce contrast agent leakage effects.

Prediction labels were associated to survival at 12 months from diagnosis (SURV12), MGMT promoter methylation, IDH mutation, Ki-67 expression and EGFR amplification. These labels were chosen as they usually provide an important prognostic information in GBM. Survival cut-off at 12 months was set based on previous studies (21–23).

*2.6 Image Processing and Radiomic Feature Extraction*

The radiomic workflow of our analysis is summarized in Figure 1. For every patient, the MRI sequences acquired were automatically co-registered with reference to MPRAGE using FMRIB's Linear Image Registration Tool (FLIRT) of FSL (24). Three regions of interest (ROIs) were manually drawn on MPRAGE and FLAIR images by a neuroradiologist (L.P., with 6 years of experience in radiology) using 3D-Slicer (25). Doubtful cases were solved as for consensus with a senior neuroradiologist (A.B, with 25 years of experience in radiology). Regions of interest (ROI) were contrast-enhancing tumor (CET), necrosis (NEC) and whole tumor including peritumoral edema (T2). A further non-enhancing tumor (NET) ROI was obtained from T2, CET and NEC ROIs as follows: T2 – (CET+NEC). Based on recent findings (26), we performed intensity non-standardness correction on our multi-institutional data by scaling each image with respect to its mean value within specific brain structure (*i.e.* NET ROI) using MATLAB R2017a environment (MATLAB. (2017). version 9.2 - R2017a). Natick, Massachusetts: The MathWorks Inc). We did not rescale the intensity range between 0 and 255 in order to prevent loss of information related to image down-sampling.

Radiomic features were extracted from NET, CET and NEC on ADC, FLAIR, MPRAGE, rCBV, T1-weigthed and T2-weighted images by using Pyradiomics package on Python 2.7 (27). In particular, 14 shape features, 18 intensity features and 75 texture features (GLCM, GLDM, GLSZM, GLRLM, NGTDM) were extracted from original and filtered (wavelet decomposition, Laplacian of Gaussian, exponential, logarithmic and gradient) images. Additionally, we included three fractal features: box counting 2D (28), box counting 3D (29) and differential box counting (30), properly adapting the code of the Pyradiomics pipeline. Only for survival prediction, along with radiomic features, we also considered the age at diagnosis. A set of 1871 features was therefore obtained for each patient and ROI-sequence combination.

*2.7 Feature selection and Classification*

The pipeline was written in Python and was implemented on Google Colab (31). Prior to any further analysis, each extracted feature distribution was standardized by removing the mean and scaling it to unit variance with Python Standard Scaler package. Feature selection was then performed in order to identify an ensemble of the most predictive features for each ROI-sequence combination. To this purpose, we used the Boruta algorithm, a powerful and recently introduced feature selector method, that trained a Random Forest Classifier on a duplicated dataset (composed by original and shadow features) and marked a feature as important comparing its Z-scores with that of the duplicate (32). The implementation we used in this work was boruta_py module, freely accessible from github repository (33). Due to the retrospective nature of this study, some MR sequences were not acquired for all the patients and some patients lacked full genetic testing, leading to class imbalance issues. In order to overcome this limitations in binary classification, we used Synthetic Minority Over-sampling Technique (SMOTE) approach, which oversamples data of the minority class, creating new synthesized samples from the existing ones (14,34).

To find the best parameter setting, an optimization search grid algorithm was applied on nine ML classifiers: AdaBoost (AB), Extreme Gradient Boosting (xGB), Gradient Boosting (GB), Decision Tree (DT) and Random Forest (RF), Logistic Regressor (LR), two Stacking classifiers (ST, ST_ABC), and KNeighbors (KN). AB, xGB and GB use a set of weak learners and try to boost them into a strong learner. The GB classifier appears in classification studies (14), as it works well with categorical and numerical data; we decided to compare GB performance with xGB, that is the fastest implementations of gradient boosted trees (14,35). The AB was also often used for brain tumor classification (36,37), as it works to create a powerful algorithm where instances are reweighted rather than resampled. A Decision Tree algorithm was used in AB as a weak learner. Decision Tree (DT) and Random Forest (RF) are both based upon decision tree algorithms. RF is actually a collection of DT attempting to classify a new object based on its attributes (38). The RF classifier was already used in brain tumor segmentation problems (39), for the MGMT promoter prediction model (40), for the IDH status prediction (41), and for the survival prediction (42). Logistic Regressor (LR) is one of the most used linear classifier to disentangle linear relationship between the data (14). The stacked generalization is an ensemble ML algorithm that learns how to best combine the predictions from multiple well-performing ML models. In our case one classifier was set on the best parameters from GB, RF and LR (ST) whereas the second was set on best parameters from GB, RF and AB (ST_ABC) (43). KN relies on distance in data space and is one of the simplest of all the supervised ML algorithms (18). Apart from the extreme gradient boosting classifier which was implemented in xgboost package (44), all classifiers were part of Scikit-learn package (45). Algorithms were chosen based on their known performance and extensive use in the literature.

In order to achieve the most performant and robust model, the Grid Search algorithm, as implemented in Scikit-learn package, was applied when computing 4 times K-fold cross validation (K=10). Given the unbalanced condition for all molecular predictors and in order to reach the same number of trials as for SURV12, an iterative way of K-fold cross validation was applied. This method made sure that among the possible combinations of data splitting, only those one having the number of minority class subjects at least equal to half of the number of majority class were included among the eligible reshuffles.

The Grid Search algorithm was set to look for the highest mean along with the lowest spread of accuracy. The accuracy mean and standard deviation were evaluated on 40 different splitting of training and test data. Once optimal parameters were identified, model performances were also assessed in terms of AUC-ROC curve (17,46). AUC-ROC curves were also useful when comparing classifiers as they show the trade-off between false positive and true positive rates in the classification (47).

## 3. Results

The study included 156 adult patients (mean age = 62 y, range = 35-83 y) with confirmed diagnosis of GBM: 121 patients were acquired at Institution 1 and 35 patients at Institution 2. Descriptive statistic performed on genetic variables revealed an odds ratio for SURV12 (0/1) of 0.607, 1.186, 0.911 and 5.6 when looking respectively at Ki-67, MGMT, IDH and EGFR. Nine ML classifiers were compared in the present study. Number of patients and label distributions for label-sequence combination are summarized in Table 1. For each label (SURV12, MGMT, IDH, KI67 and EGFR), we identified the best ROI-sequence combination in terms of prediction accuracy, reported for all the tested classifiers. For those labels suffering from class imbalance issues, SMOTE was always used. Feature selection produced multiple radiomic signatures, ordered by importance for the predicted label. According to the employed threshold for the Boruta algorithm, radiomic signatures were composed by 20 features. All box-plots of the best results and the best 15 features for every signature are displayed in supplementary materials.

As reported in Table 2, features extracted from ADC and FLAIR within the NET ROI showed high performance in discriminating SURV12, reaching maximum accuracy values respectively with AB (73.3%) and xGB (72%) classifiers, with also ROC curves mean values equal to 73.2% and 72.4% respectively (Figure 2a-2b). High results were also obtained from NEC ROI on T2 images, with maximum accuracy and ROC-AUC values achieved by xGB (Acc=74.5%; ROC-AUC=74.2%), (Figure 2c).

Best results for MGMT prediction (Table 3) were obtained from CET ROI on FLAIR images by using the AB classifier (71,7% and ROC-AUC of 70,6%), (Figure 3).

IDH prediction task showed the best performance in our dataset (Table 4). Highest accuracy was achieved from NET ROI on rCBV with AB classifier (accuracy equal to 88% and ROC AUC equal to 87.5% Figure 4a) and on T1 with ST classifier (accuracy equal to 85% and ROC AUC equal to 85%, Figure 4b). Also, T2-based features from CET ROI achieved an accuracy of 84% and ROC AUC of 82,5% with AB classifier (Figure 4c), and T2-based features from NEC ROI achieved an accuracy of 80,3% and ROC AUC of 79% with AB classifier (Figure 4d).

The Ki-67 molecular parameter provided excellent results from ADC sequence and CET ROI (Table 5). AB classifier achieved the highest prediction accuracy value (86,6%) and ROC AUC value (71,7%) (Figure 5).

Prediction accuracy results for EGFR (Table 6) were optimal for rCBV and T2 images within CET ROI, both with AB classifier. Particularly, rCBV demonstrated the highest accuracy 81% and ROC AUC 78,3% (Figure 6a), while T2 sequence achieved accuracy 81,6% and ROC AUC equal to 74% (Figure 6b).

## 4. Discussion

AI has proven to be an accurate tool in predicting survival and molecular profile of GBM. However, high variability in results across studies and lack of standardization are limiting its use in clinical practice. In the present study we aimed to shed light on the best ROI-sequence combination for prediction of clinically relevant variables by examining nine ML classification models and comparing them in terms of overall accuracy and ROC-AUC. According to our results, the AB classifier produced the best classification performance overall, with accuracy of 73.3%, 71,7%, 88%, 86,6% and 81,6% for SURV12, MGMT, IDH, Ki-67 and EGFR respectively, while the LR and KN classifiers always produced suboptimal prediction performances. These results are in line with the literature comparing boosting and logistic regression-based classifiers (48).

Focusing on survival prediction in GBM, previous studies showed highly heterogeneous results (9,18,49), depending on size and source of datasets, type and number of extracted features and model parameters. In our study the best survival prediction was achieved by AB using ADC maps from NET ROI (73.3%). Also, xGB classifiers showed high performance using T2 images from NEC ROI (74.5%) or FLAIR images from NET ROI (72%), but with higher spread of accuracy (Table 2). High performance of NET-dependent features to predict patient survival is supported by the literature. MRI peritumoral T2 hyperintensity is a common finding in GBM and is considered a combination of infiltrating tumor cells and vasogenic edema, (50), whose extension correlates with poor prognosis (51). After surgical resection, recurrence occurs more frequently along the resection margins, due to populations of malignant cells interspersed in the NET (52). Recent researches demonstrated that peritumoral MRI textural features from FLAIR and T2 images were predictive of survival as compared to features from enhancing tumor, necrotic regions and known clinical factors (53,54). High survival predictivity for ADC on NET is coherent with pathology studies demonstrating the inverse correlation between ADC values, T2-FLAIR signal intensity and tissue cellularity (55,56). In fact, tissue cellularity as measured by ADC values can differentiate between vasogenic edema and malignant tumoral tissue within the NET, possibly recognizing patients at higher risk for recurrence (57), which deeply affects OS. If confirmed by future studies, radiomic features may represent a non-invasive tool to pre-operatively stratify GBM patients based on survival, providing useful indicators to guide peritumoral resection radicality, especially for young patients (58). Good survival predictivity on NEC ROI is also supported by previous literature. Chaddad et al. reported that shape features, particularly those extracted from necrotic regions, can be used to effectively predict OS of GBM patients (59). Furthermore, our best performing feature for survival prediction on NEC was related to fractal dimension (Figure S2c), a measure of shape complexity that has rarely been employed in radiomic studies, but demonstrated interesting correlations with patient survival (60).

Preoperative prediction of MGMT promoter methylation and IDH mutation represents a crucial objective for radiomic studies due to their pivotal role in patient outcome (4,5). On conventional and advanced MR, MGMT methylated GBM may show mixed nodular enhancement, limited edema, lower rCBV, increased Ktrans, and higher ADC minimum values (61,62). IDH mutant tumors usually show less enhancement, less blood flow on perfusion weighted images, higher mean diffusion values, smaller size and frontal lobe location (12). Many studies tried to correlate these characteristics with MGMT and IDH status, reporting conflicting results (61).

Textural features demonstrated higher accuracy for MGMT promoter methylation prediction, achieving best performance with FLAIR features from CET (71,7%, AB classifier) (Figure S3 and S4). These results are coherent with other reports (63), and confirm that textural features outperform morphological and intensity features in MGMT status prediction (8). Another recent study from Sasaki et al. reported accuracy of 67% for MGMT prediction with textural features (64). A possible explanation for the performance discrepancy is the choice of the classification algorithm: prediction accuracy has great variability depending on the selected model (Table 3), with higher performance for meta-algorithms.

Regarding IDH mutation, our AB classifier achieved an accuracy of 88% with rCBV-derived first order features (median, skewness) from NET (Figure S6a), outperforming most of previous models (12,13). Kieckegereder et al. demonstrated that *IDH* mutation status is associated with a specific hypoxia/angiogenesis transcriptome signature predictable through perfusion MR (65). Our results seem to confirm a role for perfusion-based analysis in discriminating IDH mutation, reflecting the known correlation with hypoxia inducible factor (HIF) and neoangiogenesis (66). Also, textural features achieved optimal results in the prediction of IDH mutation based on T1 images from NET (85%, ST classifier) and T2 images from CET (84%, AB classifier). The accumulation of D-2HG derived from IDH mutation induces epigenetic changes that lead to abnormal gene expression and impaired cellular differentiation, possibly contributing to intra-tumoral heterogeneity. Hsieh et al. demonstrated that textural features can differentiate IDH mutation with 85% accuracy in 39 patients with GBM. The Authors performed tailored biopsies demonstrating an agreement between prediction results and biopsy-proven pathology of 0.60 (67). Shape features of tumor necrosis demonstrated good accuracy for IDH mutation prediction in our model (80,3%, AB classifier) (Figure S6d). Such result may partly explain the relation between necrosis shape and survival as shown by our results and by other Authors (59,60).

Ki-67 is a nuclear protein expressed by cells entering the mitotic cycle. In gliomas, the expression of Ki-67 is roughly proportional to the histologic grade, representing a proliferative index with prognostic correlation (68). Radiomic models predictive of Ki-67 expression have not been investigated before in the literature. In our analysis we achieved an accuracy of 86,6% for predicting Ki-67 expression through the AB. Intriguingly, best performing features were texture-based parameters extracted from the solid tumor (CET ROI) on ADC maps (Figure S8). These results perfectly agree with the role of Ki-67 as proliferative index in GBM, being ADC an MR surrogate of cellularity (55,56).

EGFR is a transmembrane tyrosine-kinase receptor for different growth factors, whose activation leads to DNA synthesis and cellular proliferation (69). Amplification of EGFR (especially EGFRvIII) is a common somatic mutation in GBM (5). Despite failure of initial attempts of targeting EGFR for therapy, the receptor remains of value for possible future treatments (69). In our results, EGFR showed best prediction performance with ST and AB classifiers. Particularly, rCBV features achieved a performance of 81% with AB classifier and T2 features achieved a performance of 81,6% with AB classifier on CET ROI. Highest scoring features were median intensity values for rCBV and textural features for T2 (Figure S10a and b). These results are supported by previous evidence. Hu et al. demonstrated a link between EGFR amplification and rCBV textural features, with correlation to microvessel volume and angiogenesis on tumor biopsies (70). Similarly, T2 textural features were shown to correlate to EGFR amplification (70).

Our study is not the first to compare different classifiers in GBM, although with different methodologies and results. Osman performed GBM patients' stratification based on conventional MRI sequences with several classifiers. Combining 9 selected radiomic features with clinical factors (*e.g.*

age and resection status), even the best prediction accuracy of the ensemble learning classifier appeared low (less than 60%), possibly due to the multi-institutional nature of the study (18). In our approach, we made use of advanced sequences and a larger number of features. Among them we also included fractal dimension-based features which have rarely been implemented in previous studies, and may helped boosting up the accuracy of our results. Further and important difference regards the use of a Boruta algorithm to reduce the features and select only those having higher importance for the model. Also Kickingereder et al. proposed to evaluate the association of multi-parametric MRI features with molecular characteristics (*e.g.* global DNA methylation subgroup, MGMT, EGFR, etc.) in GBM patients, training different models (*e.g.* stochastic GB, RF and penalized LR). The authors found associations between established MRI features and molecular characteristics (prediction accuracy of 63% for EGFR with penalized LR). However, the link between them was not strong enough to enable generation of ML classification models for reliable and clinically meaningful predictions (14). In addition to a different set of predicted outcomes, this result might be due to the type and amount of imaging features used for prediction: Kickingereder et al. used 31 imaging parameters for molecular characteristic prediction, while this study extracted 1871 radiomic features from each image.

Our study had some limitations. Firstly, even though ML studies in GBM often rely on limited populations (10, 11, 21-23, 46, 59, 67, 70), our sample size (156 patients) could be considered small. Nevertheless, our dataset include clinical/genetic information (e.g. survival, MGMT, IDH, EGFR and KI67), together with radiomic data from different MRI sequences (e.g. MPRAGE, FLAIR, ADC, rCBV, T1-wiethed and T2-weighted), thus allowing us to combine information from different sources to better predict clinical and genetic variables.

Due to the retrospective nature of the study, some sequences were not acquired for all the patients (Table 1). For this reason, prediction accuracy for each label was evaluated separately on each sequence, thus limiting performance bias. Moreover, some labels were not available for all the patients, consequently the number of subjects split in train and test groups changed for each label-sequence combination. We tried to overcome this limitation by employing two well-known and effective techniques with the aim of balancing the asymmetric labels. Although undersampling of the majority class was considered a more effective approach respect to an oversampling method (71), we decided to use SMOTE for unbalacing issues. As demonstrated in other SMOTE-based studies (14,72), it could represent a suitable solution for our purposes. In order to overcome main SMOTE drawbacks (73,74 we perform ML analysis with a significant number of cross-validations.

Since we only split subjects into train and test groups, the lack of an additional validation cohort could represents a limitation of this study. To overcome this issue, we decided to report range of performance obtained applying 4 times stratified K-fold cross-validation. This approach provides a full accuracy range, which includes the results that an eventual validation test would produce.

## 5. Conclusions

In the present study we were able to predict patient OS and highly relevant molecular features of GBM from pre-operative MRI, comparing different ML classifiers to achieve best performance. AB, ST, GB and xGB classifiers showed optimal performance in prediction tasks for the studied variables. In particular, AB and xGB obtained maximum accuracy for survival (73.3% and 74.5% respectively), AB for IDH mutation (88%), MGMT promotor methylation status (71,7%) and Ki-67 expression (86,6%), and EGFR amplification (81,6%). Best performing features from our analysis shed light on possible correlations between MR and tumor histology and may set a path for GBM ML analysis standardization.

**Conflicts of Interest:** The authors declare that the research was conducted in the absence of any commercial or financial relationship that could be constructed as a potential conflict of interest.

**Author Contributions:** L.P. and A.N. contributed equally to this work. Conceptualization, Luca Pasquini and Antonio Napolitano; Data curation, Luca Pasquini, Francesco Dellepiane, Veronica Villani, Giulio Ranazzi, Maria Camilla Rossi-Espagnet and Alberto Di Napoli; Formal analysis, Martina Lucignani and Emanuela Tagliente; Investigation, Luca Pasquini, Antonio Napolitano, Martina Lucignani and Emanuela Tagliente; Methodology, Luca Pasquini, Antonio Napolitano, Andrea Romano, Alessandro Bozzao; Project administration, Luca Pasquini; Resources, Antonio Napolitano, Antonello Vidiri, Veronica Villani, Antonella Stoppacciaro and Alessandro Bozzao; Software, Luca Pasquini, Antonio Napolitano, Martina Lucignani and Emanuela Tagliente; Supervision, Maria Camilla Rossi-Espagnet, Matteo Ritrovato, Antonello Vidiri, Antonella Stoppacciaro, Andrea Romano and Alessandro Bozzao; Validation, Antonio Napolitano and Matteo Ritrovato; Visualization, Francesco Dellepiane, Maria Camilla Rossi-Espagnet and Alberto Di Napoli; Writing – original draft, Luca Pasquini, Antonio Napolitano, Martina Lucignani, Emanuela Tagliente and Giulio Ranazzi; Writing – review & editing, Luca Pasquini, Antonio Napolitano, Martina Lucignani, Emanuela Tagliente, Francesco Dellepiane, Maria Camilla Rossi-Espagnet, Matteo Ritrovato, Antonello Vidiri, Veronica Villani, Giulio Ranazzi, Antonella Stoppacciaro, Andrea Romano, Alberto Di Napoli and Alessandro Bozzao.

## TABLES

**Table 1.** Number of patients and label distributions for label-sequence combination

|  | ADC | FLAIR | MPRAGE | rCBV | T1 | T2 |
|---|---|---|---|---|---|---|
| SURV12 (0/1) | 134 (65/69) | 140 (68/72) | 138 (66/72) | 93 (45/48) | 122 (61/61) | 122 (60/62) |
| MGMT (0/1) | 110 (41/69) | 115 (43/72) | 114 (42/72) | 80 (33/47) | 100 (39/61) | 102 (39/63) |
| IDH (0/1) | 86 (71/15) | 89 (74/15) | 89 (74/15) | 60 (51/9) | 77 (63/14) | 78 (65/13) |
| KI67 (0/1) | 100 (18/82) | 106 (21/85) | 103 (22/81) | 77 (16/61) | 97 (17/80) | 94 (16/78) |
| EGFR (0/1) | 65 (21/44) | 69 (23/46) | 66 (23/43) | 49 (16/33) | 65 (22/43) | 62 (20/42) |

**Table 2.** Surv12 best results (reported as mean±standard deviation)

| ROI | SEQ |  | xGB | GB | RF | LR | ST | KN | DT | AB | ST_ABC |
|---|---|---|---|---|---|---|---|---|---|---|---|
| NET | ADC | Acc % | 71,8±10 | 68±9,7 | 67,9±6,3 | 46±5,3 | 71±9 | 61±12,3 | 60,5±12,9 | **73,3±8,9** | 66,4±13 |
| NET | ADC | Roc % | 71,8±9,7 | 67,9±10,5 | 67,6±6 | 45±4,4 | 71,3±8,6 | 61,4±12,5 | 60,2±11 | **73,2±9** | 65±12,1 |
| NET | FLAIR | Acc % | **72±13,9** | 67,5±11,4 | 71,7±7,5 | 62±13,6 | 69±12,6 | 54±15 | 60,5±13,3 | 68,9±7 | 57,8±14,2 |
| NET | FLAIR | Roc % | **72,4±14** | 67,3±10,8 | 71,7±8,1 | 62,3±13,7 | 69±12,5 | 53,9±14,8 | 59,3±13,4 | 69,8±7,9 | 59,3±13,1 |
| NEC | T2 | Acc % | **74,5±10,9** | 63,9±13,9 | 67,3±16,8 | 58,7±14,3 | 73,7±9 | 52,3±15,2 | 59,8±10,6 | 72,7±9,5 | 58,5±14,4 |
| NEC | T2 | Roc % | **74,2±10,9** | 66±11,4 | 66,2±17 | 58,8±14,4 | 73±10 | 52±14,9 | 60,7±10,6 | 72,5±9,5 | 57,4±15,4 |

**Table 3**. MGMT best results (reported as mean±standard deviation)

| ROI | SEQ | | xGB | GB | RF | LR | ST | KN | DT | AB | ST_ABC |
|---|---|---|---|---|---|---|---|---|---|---|---|
| CET | FLAIR | Acc % | 62,6± 11,8 | 67,5± 10,6 | 65,6± 11,6 | 59± 9 | 68,7± 16 | 52± 12 | 56± 12 | **71,7± 14** | 61,4± 12,7 |
| CET | FLAIR | Roc % | 64± 11 | 66± 14 | 64± 12 | 56,5± 13,6 | 65± 16 | 51± 13 | 57± 12,8 | **70,6± 16** | 58± 13 |

**Table 4**. IDH best results (reported as mean±standard deviation)

| ROI | SEQ | | xGB | GB | RF | LR | ST | KN | DT | AB | ST_ABC |
|---|---|---|---|---|---|---|---|---|---|---|---|
| NET | rCBV | Acc % | 84,2± 13,4 | 84,2± 10,5 | 80,4± 13,4 | 80,8± 14,7 | 86,2± 10,4 | 65,4± 16,8 | 78,3± 14 | **88± 11** | 81,7± 13,8 |
| NET | rCBV | Roc % | 83,3± 13,9 | 81,7± 12,2 | 74,2± 16,2 | 79,2± 12,2 | 85,4± 11,2 | 66,2± 18,8 | 76,7± 15,2 | **87,5± 11,6** | 82,9± 8,7 |
| NET | T1 | Acc % | 79,6± 15,9 | 81,1± 14,2 | 79,9± 11,3 | 68,7± 10,9 | **85± 13** | 65± 22 | 74± 13 | 85,0± 15,6 | 79,4± 11,2 |
| NET | T1 | Roc % | 78,9± 14,5 | 79,8± 14,1 | 79,8± 11,7 | 67,3± 10,5 | **85± 13** | 65,7± 18,7 | 75,3± 13,1 | 84,8± 14,7 | 78,9± 14 |
| CET | T2 | Acc % | 82,2± 12,1 | 79,4± 12 | 81± 13 | 58± 12,6 | 81,2± 10 | 68,4± 13 | 76± 14 | **84± 10,4** | 83,7± 11,2 |
| CET | T2 | Roc % | 81,6± 11,8 | 81,6± 12,5 | 81,9± 13,1 | 59,7± 14,1 | 82,2± 11,5 | 70± 16,5 | 76,2± 13,6 | **82,5± 11,8** | 82,2± 11,1 |
| NEC | T2 | Acc % | 78,8± 8 | 78,8± 9 | 78± 11 | 68,8± 12,3 | 80± 9 | 69,8± 15,5 | 76,4± 13 | **80,3± 9,4** | 78,7± 10 |
| NEC | T2 | Roc % | 77,5± 10,3 | 78,2± 8,6 | 77,1± 10,4 | 71,2± 12,1 | 79,9± 10,6 | 69,2± 14,7 | 77± 14 | **79± 8,6** | 73± 8 |

**Table 5**. KI67 best results (reported as mean±standard deviation)

| ROI | SEQ | | xGB | GB | RF | LR | ST | KN | DT | AB | ST_ABC |
|---|---|---|---|---|---|---|---|---|---|---|---|
| CET | ADC | Acc % | 81,4± 7,3 | 80,8± 9,7 | 83,4± 9,4 | 64,7± 13,8 | 82,8± 7,6 | 66± 10 | 77,5± 10,7 | **86,6± 11,8** | 81,3± 7,7 |
| CET | ADC | Roc % | 60,8± 14,4 | 64,7± 15,8 | 66,2± 19,2 | 50,9± 18,2 | 65± 16 | 59,9± 18,2 | 77,5± 63,8 | **71,7± 20,7** | 63,3± 16,3 |

**Table 6**. EGFR best results (reported as mean±standard deviation)

| ROI | SEQ | | xGB | GB | RF | LR | ST | KN | DT | AB | ST_ABC |
|---|---|---|---|---|---|---|---|---|---|---|---|
| CET | rCBV | Acc % | 73,9± 13,7 | 78,6± 15,5 | 72,5± 15,7 | 67,2± 16,8 | 76,4± 15,5 | 60,7± 23,3 | 67,2± 20,7 | **81± 15** | 64± 20 |
| CET | rCBV | Roc % | 61,7± 20,4 | 70,5± 19 | 63,5± 20,7 | 63,2± 22,7 | 65,6± 19,7 | 64± 22 | 61± 22 | **78,3± 17,4** | 63± 20 |
| CET | T2 | Acc % | 72,9± 16,3 | 74,5± 16,3 | 74,9± 15,3 | 60,2± 18,4 | 78± 15,8 | 61± 19 | 61± 19 | **81,6± 13,7** | 72± 16 |
| CET | T2 | Roc % | 68,4± 19,3 | 67,4± 14,6 | 76,2± 14,6 | 65,6± 16,3 | 71,8± 21,6 | 60± 19,6 | 55,5± 20,4 | **74± 18** | 66± 21 |

**FIGURES**

**Figure 1**. Radiomic workflow followed in the present study

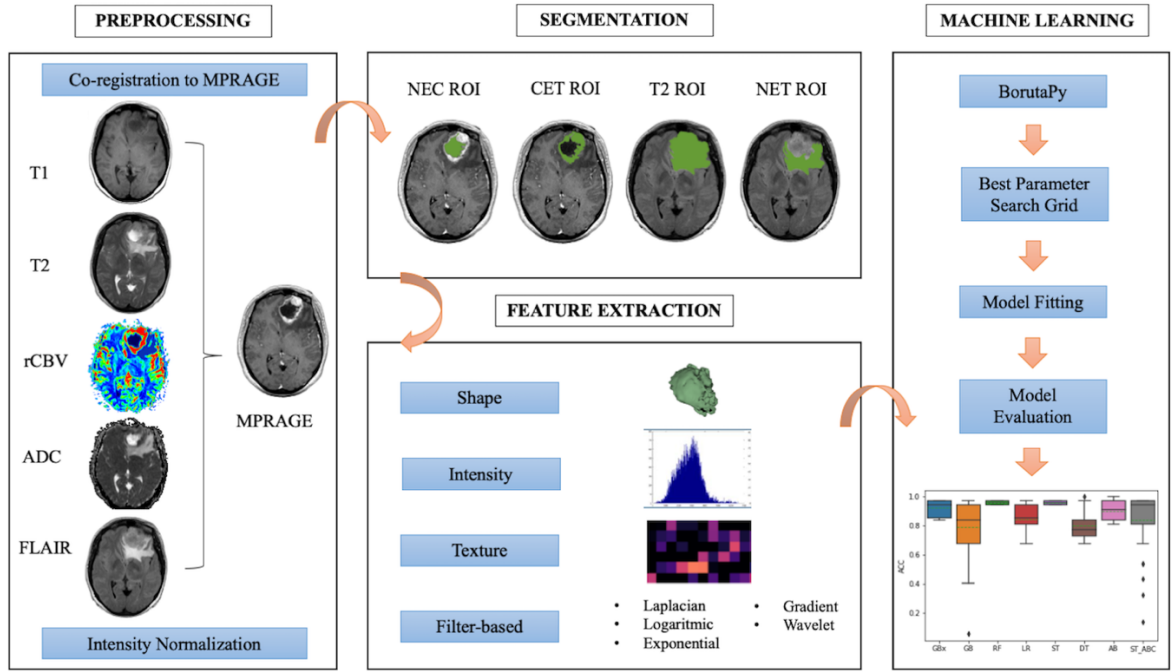

**Figure 2**. Best ROC curves for Surv12 prediction: (a) AB classifier with ADC sequence on NET ROI; (b) xGB classifier with FLAIR sequence on NET ROI; (c) xGB classifier with T2 sequence on NEC ROI

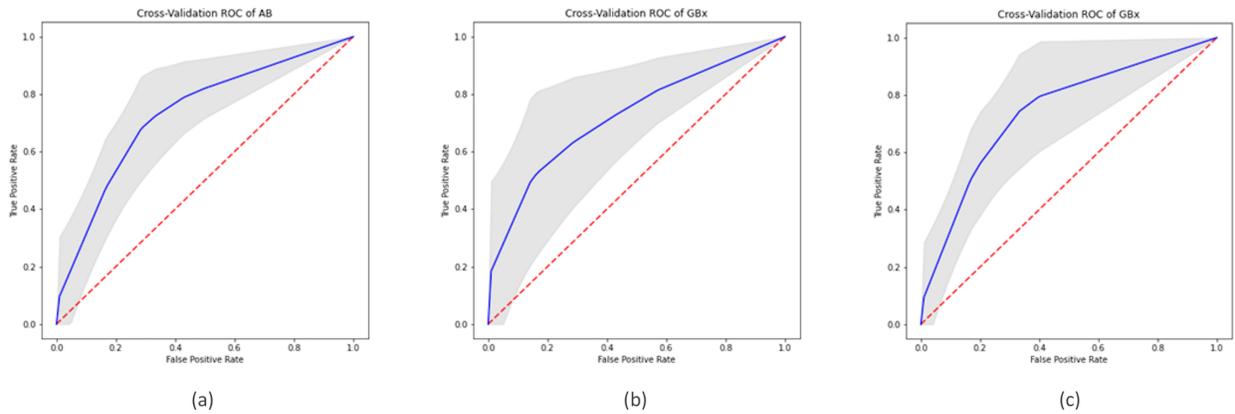

(a)  (b)  (c)

**Figure 3**. Best ROC curve for MGMT prediction: AB classifier with FLAIR sequence on CET ROI

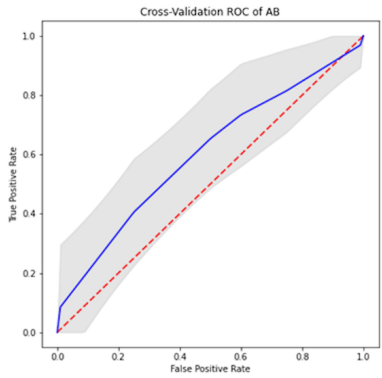

**Figure 4**. Best ROC curves for IDH prediction: (a) AB classifier with rCBV sequence on NET ROI; (b) ST classifier with T1 sequence on NET ROI; (c) AB classifier with T2 sequence on CET ROI; (d) AB classifier with T2 sequence on NEC ROI

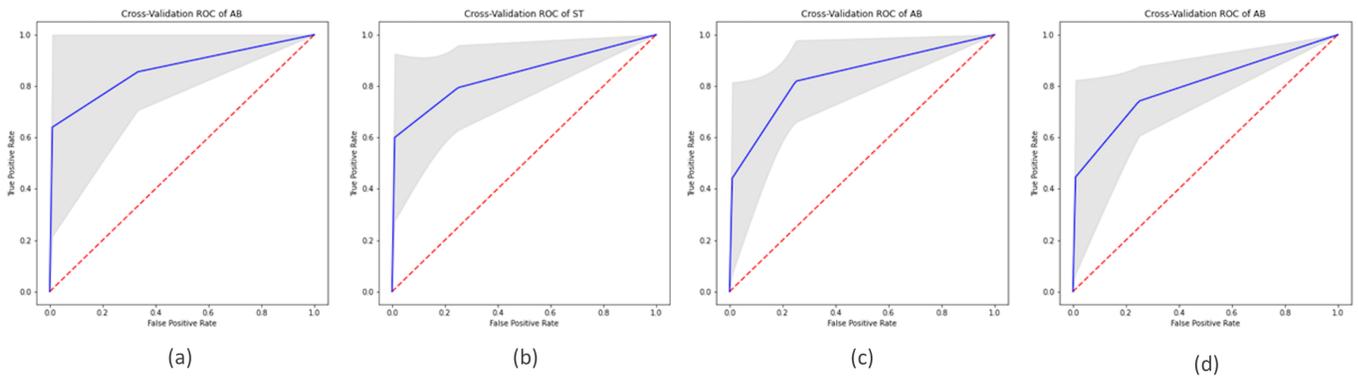

**Figure 5**. Best ROC curve for KI67 prediction: AB classifier with ADC sequence on CET ROI

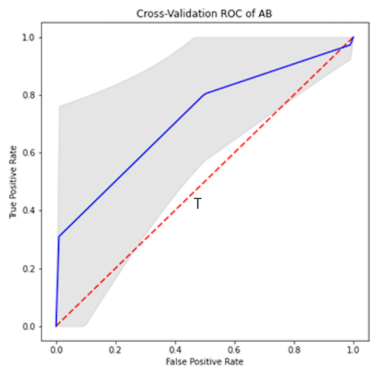

**Figure 6**. Best ROC curves for EGFR prediction: (a) ST classifier with rCBV sequence on CET ROI; (b) AB classifier with T2 sequence on CET ROI

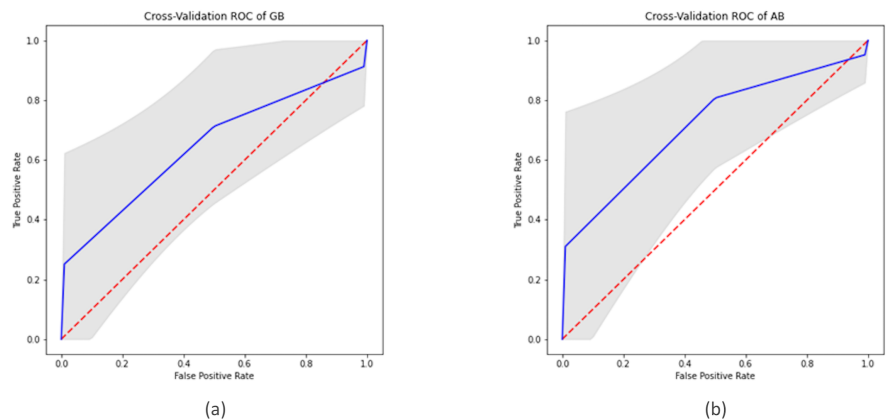